\PassOptionsToPackage{unicode}{hyperref}
\PassOptionsToPackage{hyphens}{url}
\PassOptionsToPackage{dvipsnames,svgnames,x11names}{xcolor}
\documentclass[
]{article}
\usepackage{xcolor}
\usepackage{amsmath,amssymb}
\usepackage[T1]{fontenc}
\usepackage[utf8]{inputenc}
\usepackage{textcomp} % provide euro and other symbols

\usepackage{lmodern}
\makeatletter
\@ifundefined{KOMAClassName}{% if non-KOMA class
  \IfFileExists{parskip.sty}{%
    \usepackage{parskip}
  }{% else
    \setlength{\parindent}{0pt}
    \setlength{\parskip}{6pt plus 2pt minus 1pt}}
}{% if KOMA class
  \KOMAoptions{parskip=half}}
\makeatother
\makeatletter
\ifx\paragraph\undefined\else
  \let\oldparagraph\paragraph
  \renewcommand{\paragraph}{
    \@ifstar
      \xxxParagraphStar
      \xxxParagraphNoStar
  }
  \newcommand{\xxxParagraphStar}[1]{\oldparagraph*{#1}\mbox{}}
  \newcommand{\xxxParagraphNoStar}[1]{\oldparagraph{#1}\mbox{}}
\fi
\ifx\subparagraph\undefined\else
  \let\oldsubparagraph\subparagraph
  \renewcommand{\subparagraph}{
    \@ifstar
      \xxxSubParagraphStar
      \xxxSubParagraphNoStar
  }
  \newcommand{\xxxSubParagraphStar}[1]{\oldsubparagraph*{#1}\mbox{}}
  \newcommand{\xxxSubParagraphNoStar}[1]{\oldsubparagraph{#1}\mbox{}}
\fi
\makeatother

\usepackage{longtable,booktabs,array}
\usepackage{calc} % for calculating minipage widths
\usepackage{etoolbox}
\makeatletter
\patchcmd\longtable{\par}{\if@noskipsec\mbox{}\fi\par}{}{}
\makeatother
\IfFileExists{footnotehyper.sty}{\usepackage{footnotehyper}}{\usepackage{footnote}}
\makesavenoteenv{longtable}
\usepackage{graphicx}
\makeatletter
\newsavebox\pandoc@box
\newcommand*\pandocbounded[1]{% scales image to fit in text height/width
  \sbox\pandoc@box{#1}%
  \Gscale@div\@tempa{\textheight}{\dimexpr\ht\pandoc@box+\dp\pandoc@box\relax}%
  \Gscale@div\@tempb{\linewidth}{\wd\pandoc@box}%
  \ifdim\@tempb\p@<\@tempa\p@\let\@tempa\@tempb\fi% select the smaller of both
  \ifdim\@tempa\p@<\p@\scalebox{\@tempa}{\usebox\pandoc@box}%
  \else\usebox{\pandoc@box}%
  \fi%
}
\def\fps@figure{htbp}
\makeatother

\NewDocumentCommand\citeproctext{}{}

\makeatletter
 \let\@cite@ofmt\@firstofone
 \def\@biblabel#1{}
 \def\@cite#1#2{{#1\if@tempswa , #2\fi}}
\makeatother
\newlength{\cslhangindent}
\newlength{\csllabelwidth}
\newenvironment{CSLReferences}[2] % #1 hanging-indent, #2 entry-spacing
 {\begin{list}{}{%
  \setlength{\itemindent}{0pt}
  \setlength{\leftmargin}{0pt}
  \setlength{\parsep}{0pt}
  \ifodd #1
   \setlength{\leftmargin}{\cslhangindent}
   \setlength{\itemindent}{-1\cslhangindent}
  \fi
  \setlength{\itemsep}{#2\baselineskip}}}
 {\end{list}}
\usepackage{calc}

\newcommand{\CSLLeftMargin}[1]{\parbox[t]{\csllabelwidth}{\strut#1\strut}}
\newcommand{\CSLRightInline}[1]{\parbox[t]{\linewidth - \csllabelwidth}{\strut#1\strut}}

\usepackage{arxiv}
\usepackage{orcidlink}
\usepackage{amsmath}
\usepackage[T1]{fontenc}
\makeatletter
\@ifpackageloaded{caption}{}{\usepackage{caption}}
\AtBeginDocument{%
\ifdefined\contentsname
  \renewcommand*\contentsname{Table of contents}
\else
  \newcommand\contentsname{Table of contents}
\fi
\ifdefined\listfigurename
  \renewcommand*\listfigurename{List of Figures}
\else
  \newcommand\listfigurename{List of Figures}
\fi
\ifdefined\listtablename
  \renewcommand*\listtablename{List of Tables}
\else
  \newcommand\listtablename{List of Tables}
\fi
\ifdefined\figurename
  \renewcommand*\figurename{Figure}
\else
  \newcommand\figurename{Figure}
\fi
\ifdefined\tablename
  \renewcommand*\tablename{Table}
\else
  \newcommand\tablename{Table}
\fi
}
\@ifpackageloaded{float}{}{\usepackage{float}}
\floatstyle{ruled}
\@ifundefined{c@chapter}{\newfloat{codelisting}{h}{lop}}{\newfloat{codelisting}{h}{lop}[chapter]}
\floatname{codelisting}{Listing}

\makeatother
\makeatletter
\@ifpackageloaded{caption}{}{\usepackage{caption}}
\@ifpackageloaded{subcaption}{}{\usepackage{subcaption}}
\makeatother
\usepackage{bookmark}
\IfFileExists{xurl.sty}{\usepackage{xurl}}{} % add URL line breaks if available
\hypersetup{
  pdftitle={PRITES: An integrative framework for investigating and assessing web-scraped HTTP-response datasets for research applications},
  pdfauthor={Cynthia A. Huang; Tina Lam},
  pdfkeywords={web-scraping, web crawling, big data, application
programming interface, API},
  colorlinks=true,
  linkcolor={blue},
  filecolor={Maroon},
  citecolor={Blue},
  urlcolor={Blue},
  pdfcreator={LaTeX via pandoc}}

\newcommand{\runninghead}{A Preprint }
\title{PRITES: An integrative framework for investigating and assessing
web-scraped HTTP-response datasets for research applications}
\def\asep{\\\\\\ } % default: all authors on same column
\author{\textbf{Cynthia A. Huang}\\\\Department of Econometrics and
Business Statistics, Monash University\\\\\asep\textbf{Tina
Lam}\\\\Monash Addiction Research Centre, Monash
University\\\\\href{mailto:tina.lam@monash.edu}{tina.lam@monash.edu}}
\date{}
\begin{document}
\maketitle
\begin{abstract}
The ability to programmatically retrieve vast quantities of data from
online sources has given rise to increasing usage of web-scraped
datasets for various purposes across government, industry and academia.
Contemporaneously, there has also been growing discussion about the
statistical qualities and limitations of collecting from online data
sources and analysing web-scraped datasets. However, literature on
web-scraping is distributed across computer science, statistical
methodology and application domains, with distinct and occasionally
conflicting definitions of web-scraping and conceptualisations of
web-scraped data quality. This work synthesises technical and
statistical concepts, best practices and insights across these relevant
disciplines to inform documentation during web-scraping processes, and
quality assessment of the resultant web-scraped datasets.

We propose an integrated framework to cover multiple processes during
the creation of web-scraped datasets including `Plan', `Retrieve',
`Investigate', `Transform', `Evaluate' and `Summarise' (PRITES). The
framework groups related quality factors which should be monitored
during the collection of new web-scraped data, and/or investigated when
assessing potential applications of existing web-scraped datasets. We
connect each stage to existing discussions of technical and statistical
challenges in collecting and analysing web-scraped data. We then apply
the framework to describe related work by the co-authors to adapt
web-scraped retail prices for alcoholic beverages collected by an
industry data partner into analysis-ready datasets for public health
policy research. The case study illustrates how the framework supports
accurate and comprehensive scientific reporting of studies using
web-scraped datasets.
\end{abstract}
{\bfseries \emph Keywords}
\def\sep{\textbullet\ }
web-scraping, web crawling, big data, application programming interface,
API \sep 
web-scraping, web crawling, big data, application programming interface,
API

\section{Introduction}\label{sec-intro}

Frameworks and tools for understanding the statistical quality of data
collected from public internet sources are increasingly important given
the ubiquity and popularity of such data. Web-scraping is often promoted
as a low-cost and `simple' alternative data collection method to more
traditional labour-intensive methods such as survey sampling. Datasets
collected through the use of web-scraping are increasingly common inputs
to data analyses across many research fields (e.g. {[}1{]}, {[}2{]} and
{[}3{]}). However, it has been widely acknowledged that unlike
well-designed survey datasets with clear sampling frames, web-scraped
datasets often violate traditional assumptions of randomness and
representativeness. These assumptions are needed to justify the usage of
statistical inference methods based on large-sample asymptotics, and
failing to address violations of these assumptions can lead to invalid
conclusions as discussed in {[}4{]}, {[}5{]} and {[}6{]}. Nevertheless,
with careful consideration of the data itself and the analysis context,
web-scraped data can be valuable sources for gaining insight into
important and timely research questions.

Researchers assessing the suitability of web-scraped datasets for
particular research questions have to consider factors such as how data
were collected and from which websites, if and how statistical units and
derived variables were constructed from the collected data, exactly what
the processed data are representative of, and known limitations of the
dataset. Meng {[}7{]} proposes the terms `data minding' and `data
confessions' to refer to the practice of documenting and disclosing
information relevant for the assessment of datasets. However, as we will
discuss, investigating and documenting relevant factors for web-scraped
data requires combining applying expertise from survey methodology, data
preprocessing and provenance, and web-technologies into domain-specific
application contexts. There is a need for integrated guidance on how to
document and communicate technical and statistical qualities of
web-scraped datasets, as well as how to assess the suitability between a
given dataset and particular lines of research inquiry. To this end,
this paper aims to contribute to the development of such guidance by
synthesising insights from data engineering, survey methodology and
information science to offer a multi-layer framework for identifying
data quality factors, and offering an example case study which applies
the framework to assess and adapt a commercially collected web-scraped
dataset for research use.

Our multi-stage framework organises six processes associated with the
creation of web-scraped datasets into four stages -- (1) `Plan', (2)
`Retrieve', (3) `Investigate, Transform and Evaluate', and (4)
`Summarise' (PRITES). The Planning stage involves defining the target
population and collection plan. The Retrieval stage focuses on the
technical execution of the web-scraping processes. The third stage
consists of three related, often iterative and/or nested processes.
Investigation aims to understand what was retrieved, compare the
retrieved objects with expectations, and characterise factors specific
to data sources and the collection process that might affect dataset
quality. Transformation involves documenting and implementing data
wrangling operations including parsing, cleaning, and integration with
other data, and directly impacts subsequent downstream analysis.
Evaluation involves assessing the transformed data, along with insights
gained through investigation, against the intended analysis and research
context. Depending on the outcome of the evaluation, users might conduct
further investigation or transformation, or move on to the final stage.
For example, a common workflow involves the investigation and
identification of available variables, extraction (i.e.~transformation)
of those variables, and then investigating the completeness of values
within a given variable before conducting more transformation. The final
`Summarise' stage focuses on clear and reproducible documentation of the
properties and suitable applications of the final web-scraped dataset.

The included case study applies the framework to relate work on
preparing commercially collected web-scraped data for research purposes.
We illustrate how a clear understanding of the strengths and weaknesses
of web-scraping as a data collection method, regular discussions with
our data provider, and targeted data validation and investigation
facilitated iterative refinement and scoping of research questions. The
framework supported the assessment of what questions the available data
could feasibly provide insight on. Such assessments not only decrease
the need for complex methodological fixes and contrived justifications
in downstream analysis, but also illuminate opportunities for
supplementary data collection and augmentation to improve data quality.

\section{Background and Motivation}\label{sec-background}

Although web-scraped data is often described as a cheap and easy
alternative to traditionally collected datasets across various domains,
the technical complexity of web-scraping processes is highly variable,
as is the availability and accessibility of internet data sources.
Variability gives rise to both technical and statistical challenges in
the collection and usage of web data. The scale, variety and sheer
magnitude of data generating and sharing processes, entities, and
organisations on the World Wide Web also give rise to opportunities and
challenges for multi-source inference on combinations of data from
internet sources. Discussions about the empirical validity of
web-scraped data often appeal to established notions from survey
methodology, such as non-response and sampling bias, with only cursory
consideration of technical implementation details. By contrast, research
on web-scraping tools and workflows generally focuses on the successful
retrieval of web content and parsing of that content into structured
data for downstream manipulation. The focus on retrieval leads to
greater emphasis on effective and user-friendly interfaces to access
internet data sources, with minimal consideration of complex data
quality issues. Existing concepts of `paradata', `metadata', and
`substantive data' from survey methodology (c.f. {[}8{]} and {[}9{]}),
as well as tools for data and workflow provenance, offer some foundation
for integrating these two perspectives. However, to the authors'
knowledge, there has not been any specific guidance for assessing and
reporting web-scraped data quality that spans both technical and
statistical considerations.

In order to integrate these complementary perspectives, we first clarify
a number of terms and concepts relevant to web-scraping. The
distribution of web-scraping literature across multiple disciplines and
independent innovations in accessing and collecting data from internet
sources has led to occasional conflicts in nomenclature. The naming of
inputs, data sources, collection methods, and processes, as well as the
retrieved data, is highly varied. The varied definitions reflect subtle
but important distinctions between web-scraping as a set of technologies
for data retrieval and as a statistical data collection method.

\subsection{Defining web-scraping}\label{sec-background-defining}

The definition of `web-scraping' varies in precision, scope and
technicality depending on context. The Cambridge Dictionary defines
{\emph{web-scraping}} as ``the activity of taking information from a
website or computer screen and putting it into an ordered document on a
computer'' {[}10{]}. This task-focused definition encompasses both the
retrieval of web data and the structuring of that data, with minimal
consideration of statistical qualities of such data. By contrast,
discussions of web-scraping in empirical social science focus
predominantly on the opportunity to collect data from novel sources and
data generating processes. For example, web-scraping is defined as ``the
automated process of accessing websites and downloading their content''
in {[}3{]}, {[}11{]}, and {[}12{]} . Similarly, discussions of
web-scraping as an alternative or complement to traditional survey
methodologies may not even use or define the term `web-scraping'.
Instead, terms like `big data' (e.g., in {[}1{]}) and `internet data
sources' (e.g., in {[}13{]}) are used to discuss potential statistical
issues with collecting data from the internet, irrespective of the exact
collection processes and technologies used.

\subsubsection{Web technologies}\label{sec-background-webTech}

Although a detailed explanation of all the physical and digital
infrastructure that enables the World Wide Web is beyond the scope of
this paper, it is useful to have a working understanding of the internet
as an ecosystem of digital entities and processes which produce data
objects and the technologies that facilitate programmatic attempts to
collect such objects.

{\emph{Webpages}} are rendered documents intended for human viewing on a
{\emph{client}} machine, retrieved from a {\emph{web server}}.
{\emph{Websites}} are navigable collections of webpages, some of which
might be publicly accessible, while others might require
{\emph{authentication}} to view. Instructions for how a document should
look and what content it should show are written using client-side web
technologies such as {\emph{HTML}}, {\emph{CSS}}, {\emph{Javascript}}.
{\emph{Rendering}} is the process of following those instructions to
retrieve and combine content such as text, images, and videos into a
document displayed in a {\emph{web browser}}. Webpages can contain both
{\emph{static}} components, which are always the same every time you
visit, and {\emph{dynamic}} components, which retrieve content during
the rendering process to display. {\emph{Dynamic}} components can be
used to display up-to-date or personalised data on webpages without
needing to modify the document itself. Data for dynamic components are
usually stored in databases and {\emph{queryable}} via
{\emph{Application Programming Interfaces (APIs)}}. {\emph{Web
Applications}} refer to interactive software that is accessible via a
web browser and often contain dynamic content.

\subsubsection{Online datasets vs.~web-response
datasets}\label{sec-background-webResponseDatasets}

In this work, we address the interaction between web-scraping as a data
\emph{collection} method and the quality of datasets containing
web-scraped data. We exclude from this discussion the use of web
technologies to download existing datasets, where the impact of
web-scraping could be reduced to a question of retrieval success --
i.e.~``was the entire dataset and any relevant metadata about the
dataset retrieved?''. We use the term {\textbf{online datasets}} to
refer to a structured collection of data \emph{ex-ante} conceptualised,
documented, and published for the purpose of further processing or
analysis retrieved from a web server. Furthermore, we refer to anything
returned by a \emph{web server} in response to requests for dynamic or
static content as {\textbf{web-response objects}}.

The retrieval success of an existing {\emph{online dataset}} from a web
server involves confirming whether the relevant response objects, such
as multiple compressed files for multi-file datasets, can be opened and
parsed (i.e., are uncorrupted), and when appropriately pieced together
(e.g., decompressed and merged), the resultant dataset matches expected
dataset dimensions (e.g., total table rows, pages, images etc.)
according to available dataset documentation and metadata. By contrast,
the usage of web-scraping to collect web-response objects which are then
parsed and transformed into structured datasets \emph{ex-post} should be
assessed through the lens of existing theory around data collection
methods in survey methodology and related statistical literatures.

Rather than the more commonly used but ill-defined term `web-scraped
datasets', we propose the term {\textbf{web-response dataset}} to refer
to datasets constructed fully or in part from response objects retrieved
from online sources using web technologies. The creation of web-response
datasets must involve sending requests to a web server and receiving
response objects such as HTML documents, XML/JSON data, and other
content such as images and videos\footnote{The process of obtaining
  response objects is often also referred to as `fetching', and we use
  the terms interchangeably with `retrieving'.}. We propose a new term
because some existing usages of `web-scraped data' only include
interactions with rendered HTML documents (e.g.~in {[}3{]} and
{[}14{]}). This usage excludes other methods such as API queries,
headless browsers, and HTTP requests which also return web-response
objects.

\subsubsection{Web crawling and internet data
sources}\label{sec-background-internetDataSources}

The term {\emph{web crawling}} refers to the process of identifying and
{\textbf{indexing}} internet data sources and specific resources
available for request and querying (c.f. {[}2{]}). This process
generally involves negligible collection of substantive response objects
and is most commonly exemplified by search engine indexes which only
collect web page titles and minimal content previews. In the case of
web-response datasets, the outcome of indexing is a list of potential
requests for response objects, which might take the form of a list of
{\emph{Unique Resource Links (URLs)}} or {\emph{API queries}}. These
potential requests can be arranged hierarchically according to the
target website or application they will be requested from (e.g.~multiple
product webpages from the same online web store), as well as by the
relevant owner and publisher of the requested information (e.g.~the
retailer).

We adapt the term {\textbf{internet data source}}\footnote{Beręsewicz
  {[}5{]} defines'Internet Data Source' as a ``self-selected
  (non-probabilistic) sample that is created through the Internet and
  maintained by entities external to National Statistical Institutions
  (NSIs) and administrative regulations.'' The author stress that their
  definition explicitly refers to data not collected by statistical
  institutions or public agencies but by private/commercial entities,
  and also pre-existing datasets published online by statistical
  organisations. However, this distinction unnecessarily constrains the
  scope of data sources for web-response datasets and excludes the
  possibilities such as web-response datasets constructed using data
  about NSIs collected from the NSI websites.} from Beręsewicz {[}5{]},
to refer to the entity which owns and publishes the information targeted
by the requests. A given web-response dataset could be constructed from
response objects published by a single internet data source or multiple
internet sources (e.g.~multiple retailers). The retrieval of those
objects involves the execution of the list of potential requests to
target URLs or APIs, any of which could have access restrictions
requiring some form of authentication (i.e., via some type of account
registration). Depending on the research question and technical context,
the internet data source might be distinct from the target websites or
web applications from which the response objects are retrieved. The
behaviour and characteristics of internet data sources can have a
significant influence on the quality and statistical properties of
web-response datasets. As discussed in
Section~\ref{sec-background-stats}, failure to document and account for
such factors can give rise to errors and biases in downstream
statistical analyses.

\subsection{Retrieving web-response
objects}\label{sec-background-retrieval}

In general, the technical retrieval of web-response objects requires a
variety of web programming and data engineering skills. However, there
are numerous interactive and programmatic tools, software libraries, and
tutorials for web-scraping and API access, which provide tooling and
guidance to programmers of varying backgrounds and expertise. These
existing tools tend to provide simple all-in-one interfaces which output
collected information in immediately usable formats such as tables. Such
designs fail to support provenance and statistical tasks, such as
inspecting the wrangling of the retrieved objects into the output tables
or datasets, or identifying corresponding internet data sources.
However, existing tools for tracking computational processes (i.e., data
and workflow provenance), along with the concepts of
{\emph{`paradata'}}, {\emph{`metadata'}}, and {\emph{`substantive
data'}}, may offer some practical inspiration and conceptual scaffolding
for addressing these limitations. We also note that the use of
web-scraping raises a variety of ethical and legal considerations,
though discussion of these issues is outside the scope of this paper.

\subsubsection{Existing web-scraping
tools}\label{sec-background-webscrapingTools}

There are two main types of tools for web-scraping tasks: (1) desktop
applications, which allow users to capture data from rendered webpages
via point and click interactions; and (2) programmatic libraries and
frameworks, which offer varying combinations and degrees of support for
fetching, extracting, and transforming web-response objects into
structured datasets. Both types of tools focus primarily on improving
the ease and convenience of retrieving response objects from online data
sources and creating structured datasets, with minimal to no explicit
support for tracking statistically relevant information such as response
rates or detecting potential sources of bias such as systematic data
missingness.

Furthermore, the focus on ease and convenience generally means that
retrieved response objects and associated information about the
collection process are only temporarily cached by default and then
discarded once the desired data has been extracted. For example, in the
getting started tutorials for the popular web-scraping library `rvest'
{[}15{]}, users are taught to write functions that retrieve and parse
temporary HTML documents, retaining only the targeted values and
disposing of the initial HTML response object. This narrow focus on task
completion is also reflected in the evaluations of web-scraping tools.
Persson {[}16{]} presents a survey and evaluation of state-of-the-art
web-scraping tools, with an evaluation framework based on general
software evaluation criteria and examine the following four areas:
`Performance', `Features', `Reliability', and `Ease-of-Use'.

Although combining the retrieval of response objects with data parsing
operations may be expedient from an engineering perspective, it gives
rise to significant complications for the valid usage and analysis of
web-response datasets. Data collection and processing decisions impact
the quality of web-scraped datasets through conceptually distinct
channels. However, such information can be difficult to disentangle when
using existing tools. In particular, the loss of information about the
collection process (e.g., failed retrieval attempts, properties of the
response objects) can hinder downstream efforts to select and calibrate
appropriate data analysis methods.

\subsubsection{Provenance
metadata}\label{sec-background-provenanceMetadata}

Provenance in the context of digital data generally refers to the
history and lineage of a particular piece or collection of data (i.e.,
datasets). The information required to answer provenance questions such
as what data sources were used or what transformations were applied is
referred to in aggregate by computer scientists as ``provenance
(meta)data'' {[}17{]}. Provenance data can be captured systematically
during the production of a web-response dataset (i.e., automated logging
of failed requests, recording timestamps from request and response
headers), partially inferred from production artefacts (e.g.,
web-scraping scripts, response objects, data wrangling code), or
manually documented (e.g., as narrative text alongside or within
computational notebooks).

The type and intended use of provenance data often determine the degree
to which that data can be automatically captured. For example, although
code artefacts are often sufficient for understanding how data was
extracted and transformed from retrieved response objects, they
generally lack information on the assumptions, context and reasoning
behind the code. Such assumptions can include the comparability of
values extracted from different elements of the response objects (e.g.,
justifying fallback logic for extracting values from other elements when
encountering an empty element) or the presence of structural or content
irregularities that prevent the extraction of information (e.g.,
differing text delimiters which affect the matching and extraction of
values).

Assumptions underpinning the construction and quality of web-response
datasets are often based on observations made during ad-hoc and
non-exhaustive exploration of subsets of the response objects, which are
then assumed to apply across all collected response objects.
Unfortunately, generalisations typically only hold across a small subset
of the response objects, with new cases added to the transformation code
as new parsing and transformation errors are encountered. This heuristic
trial and error approach can lead to quite complex and difficult to
interpret code artefacts, especially if conditional logic is used
extensively.

According to Ragan et. al's classification of provenance information
types {[}18{]}, `how' information about the history and changes of
web-response datasets can be considered `data provenance', whilst
`insight' and `rationale' provenance correspond more closely with `why'
questions and the assumptions discussed above. As noted by the authors,
`insight' and `rationale' provenance are difficult to capture in an
automated manner. However, it is difficult to fully assess the quality
and properties of web-response datasets without such information. As
such, we also considered the type of provenance information and
artefacts most relevant to each process and stage when designing the
PRITES framework. In particular, tools and methods for capturing
`insight' and `rationale' provenance are most relevant in the `plan',
`investigate', and `evaluate' processes, whilst existing solutions for
producing `data provenance' documentation based on code artefacts (e.g.
{[}19{]} and {[}20{]}) are most useful in the `retrieval' and
`transformation' processes.

\subsubsection{Paradata, metadata and substantive
data}\label{sec-background-paradata}

In contrast with the process-driven definitions of provenance data
discussed above, provenance information can be further divided by scope
into micro-level information about each observation unit (`paradata')
and macro-level information about a dataset (`metadata'). The former was
first conceptualised in survey methodology to refer to additional data
captured during the process of producing a survey statistic {[}8{]}. The
distinction between the two is clarified by Andersson et. al {[}21{]}:
``metadata refers to a condensed and structured description of a
resource, commonly guided by a standard agreed upon by some form of
community'', while the concept of ``paradata'' addresses the need to
``document data processing beyond what is traditionally captured in
structured metadata.'' However, as discussed by Schenk and Reuß {[}9{]},
the exact boundaries between paradata, metadata, and substantive (i.e.,
analysis-ready) data often depends on context.

In the context of web-response datasets, {\textbf{paradata}} can be
understood as information relating to the collection process of
web-response data, whilst {\textbf{metadata}} are summary attributes of
datasets and collections of {\textbf{substantive data}}. Substantive
data can refer to both ``raw'' web-response objects and rectangular
tables of ``analysis-ready'' observations, though the paradata and
metadata for each would differ. For example, paradata about the
retrieval process and raw web-response objects could be summarised into
metadata about one of multiple data sources for an analysis-ready
dataset that combines data from multiple web-scraping projects. In this
interpretation, metadata is a boundary object {[}22{]} capturing
information deemed ``relevant'' for future users or assessors of the
dataset according to the dataset publisher and/or shared community
standard. Metadata could be composed of descriptions of features of the
dataset itself (e.g., summary statistics), be constructed from paradata
(e.g., patterns of systematic non-response or missingness at the
collection stage), as well as additional annotations, notes or
documentation. By contrast, paradata is generally easier to automate
capture for, and more likely to be recorded along the same units as the
list of requests or specific internet data sources.

\subsection{Analysing web-response
datasets}\label{sec-background-analysing}

The validity of any empirical analysis depends on the quality and
relevance of the data to the analysis question, along with the
suitability and effectiveness of methods used to learn from the data.
When considering the quality of any dataset, it is important to
understand what phenomena and population the data are representative of,
and whether that matches the intended analysis questions.

In the context of web-response datasets, the information required for
that understanding includes the timing of web requests and responses,
how (and why) objects were processed and parsed, harmonised, and/or
imputed to form the final dataset. However, even when the relevant
information is available or could be collected, interpreting the
statistical implications and determining whether the data is suitable
for the intended analysis is not necessarily straightforward.
Fortunately, there is a rich literature in sampling theory and survey
methodology on identifying and characterising misalignments between
collected data and desired observations or measurements. Moreover, this
literature also provides extensive guidance on appropriate corrections
and adjustments to mitigate the impact of various sampling and
non-sampling errors on empirical validity.

\subsubsection{Related statistical
literature}\label{sec-background-stats}

According to Beręsewicz {[}13{]}, one of the earliest discussions of the
statistical properties of web-scraped data and online data sources is by
Shmueli et. al {[}6{]}, which discusses various potential sources of
sampling errors, as measured in terms of bias and variance, and
non-sampling errors for online auction data. Potential non-sampling
errors include various subtypes of selection bias, such as
under-coverage of the target population, systematic non-response,
misspecification of the target population, as well as detailing possible
forms of measurement bias, such as interviewer/experimenter effects,
non-truthful responses, errors in data recording, and poorly designed
questions in questionnaires. Extended discussion of these potential
statistical issues is beyond the scope of this paper. However, given the
common but at times misguided view of web-response data as complete
censuses of a given population, we briefly review discussions of
coverage and representativeness.

Many statisticians and empirical researchers have highlighted parallels
between {\emph{indexing}} web pages and the construction of a sampling
frame, as well as between {\emph{retrieving}} response objects and
sampling of units from those frames. Furthermore, under the assumption
that a constructed index of potential web-response requests is
exhaustive for a given internet data source, a parallel can be drawn
between the index and a full population register (i.e., a complete list
of product page URLs from a given online retailer \emph{is} a complete
register of the population of products sold by that retailer).

However, as discussed in {[}5{]}, it is not always appropriate to equate
an index of web requests with a population register. It can be difficult
to describe or characterise what the target population, observational,
and sample units are, and to confirm that the target population has been
fully registered or indexed. Foerderer {[}11{]} illustrates the
difficulty of detecting unindexed units by comparing the results of
several web-scraping runs of an online marketplace for computer games
with true data obtained directly from the owner of the marketplace. Even
with a fully indexed population, it can be difficult to retrieve all the
desired observational units. Volatility or restrictions in the
availability of data due to personalisation, rate limits, and other
difficult-to-detect factors can prevent complete censuses of a given
population. As highlighted in existing work, it is generally not
possible to determine the true population coverage percentage and
representativeness of web-scraped data without access to key information
about the internet data source such as details of personalisation
algorithms {[}23{]} and properties of the target database {[}11{]}.

Beyond issues of population representativeness, establishing the meaning
and validity of variables constructed from the retrieved response
objects requires thorough investigation and understanding of the
internet data source and may involve subjective interpretation of the
retrieved information. Landers et. al {[}14{]} propose the term `data
source theory' to refer to these assumptions about what measurement or
attribute of the population units the retrieved data represents. They
further recommend that ``web-scraping projects be explicit about
theoretical and empirical support for the assumptions they have made
about their data's existence;'' which aligns with Meng's more general
call for increased `data minding' {[}7{]}.

In addition to the above works, discussions about the statistical
properties of ``big data'', which often encompass data from online data
sources, also provide relevant guidance (e.g., {[}24{]}, {[}4{]}, and
{[}1{]}). Similarly, discussions of self-selection bias and sampling
frame coverage in web surveys are also directly applicable (e.g.,
{[}25{]}, and {[}26{]}). As noted in {[}5{]}, Internet Data Sources can
be considered ``imperfect web surveys'', whereby websites, web
applications, or databases are ``interviewed'' (i.e., queried) for
response objects. Further generalisations of these considerations lead
to broader discussions about the role of data preprocessing in the
interpretation and validity of downstream inference, estimation, and
predictions (e.g., {[}27{]}, {[}28{]}, and {[}29{]}).

\subsubsection{Existing data quality
frameworks}\label{sec-background-qualityFrameworks}

Many of the above contributions motivate their work with the observation
that insufficient attention appears to be given to sampling,
preprocessing, and other statistical issues in applied work, especially
when online data sources are involved. By contrast, our work, whilst
similarly acknowledging the need for more attention towards these
issues, does not claim any further contributions towards articulating
threats to the empirical validity of web-scraped datasets. Instead, we
attempt to operationalise and integrate these existing insights into a
practical framework and workflow for addressing these issues at the
relevant stages of collecting, assessing, and using web-scraped
datasets.

Our framework facilitates the tracking of statistical considerations
alongside the documentation and/or investigation of web-scraping
processes. In this way, our work shares similar motivations with Kenett
and Shmueli's generalised framework for assessing the Information
Quality of any given dataset {[}30{]}. We also share overlapping
concerns with domain-specific empirical guides such as Boegershausen et.
al's three-stage framework for considering ``idiosyncratic technical and
legal/ethical questions'' when collecting web-response datasets for
marketing research {[}3{]}.

\section{PRITES Framework}\label{sec-framework}

\begin{figure*}

\centering{

\pandocbounded{\includegraphics[keepaspectratio]{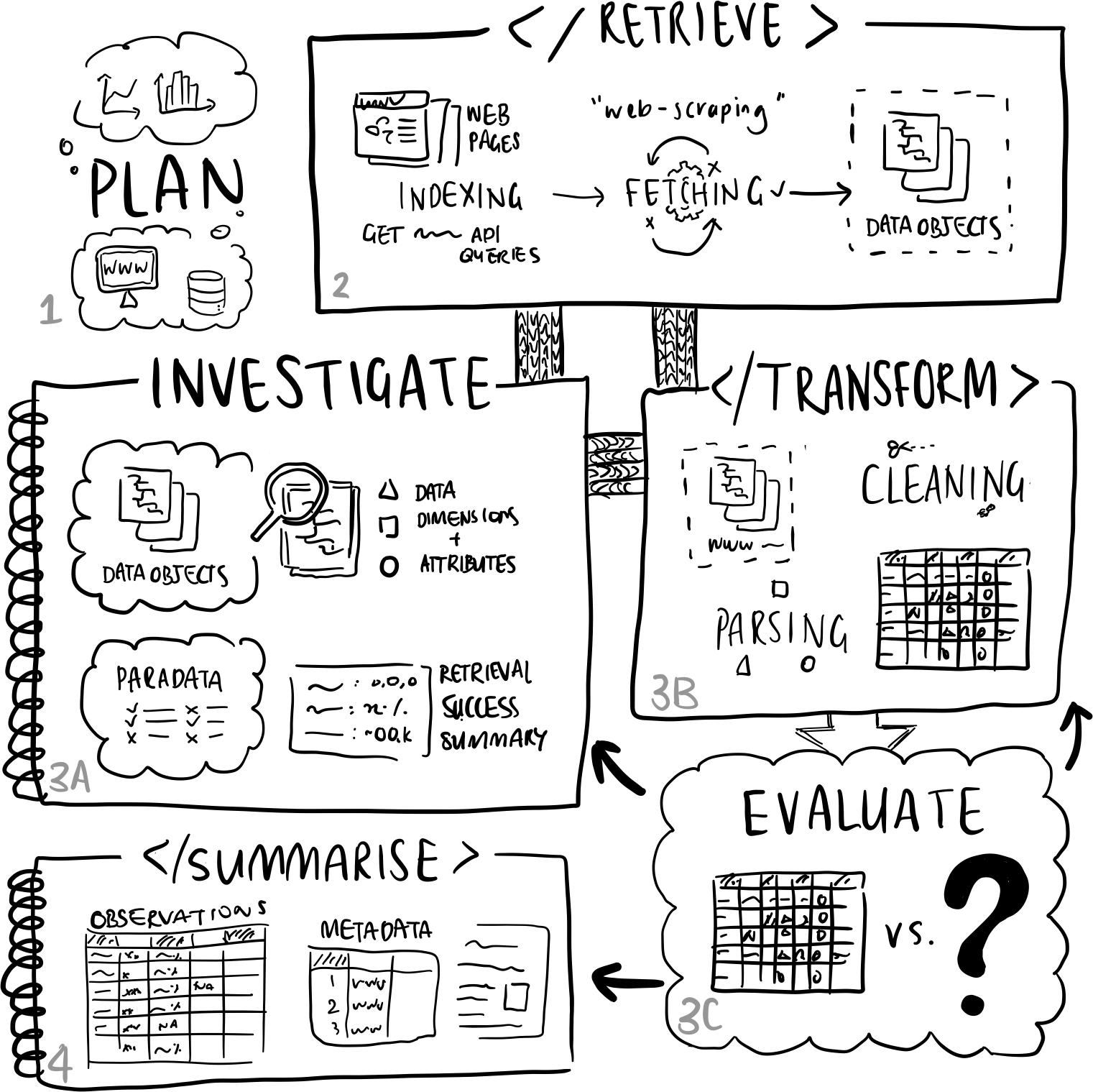}}

}

\caption{\label{fig-framework-overview}The PRITES framework structures
dataset planning and conceptualisation, retrieval, investigation,
transformation, evaluation, and summary tasks required to construct
web-response datasets into the following stages: (1) `Plan', (2)
`Retrieve', (3A) `Investigate', (3B) `Transform', (3C) `Evaluate', and
(4) `Summarise'. The style of each task box roughly corresponds to the
most relevant type of provenance and provenance documentation for that
task. The `thought bubbles' for `Plan' and `Evaluate' correspond to
rationale provenance, `spiral notebooks' for `Investigate' and
`Summarise' to insight provenance, and `code tags' for `Retrieve',
`Transform', and `Summarise' correspond to data and workflow provenance
and reproducibility.}

\end{figure*}%

Despite the number of existing frameworks and discussions, to our
knowledge, none of these existing works attempt to fully integrate
technical, conceptual, and statistical issues under a single quality
assessment framework. We offer in this work a framework of generalisable
processes and task abstractions, rather than discussing checklists of
specific issues. The proposed stages and processes correspond directly
with distinct technical or statistical expertise, such that these tasks
can be appropriately delegated rather than requiring a single researcher
to master all the requisite skills. Furthermore, we offer suggestions
for documentation and provenance artefacts that can be produced
separately, and then combined within our unifying framework to
synthesise progress and insights from the complementary technical,
statistical, and domain expertise required to construct and evaluate
web-response datasets.

\subsection{Framework overview and
goals}\label{framework-overview-and-goals}

The PRITES framework is designed to facilitate the identification and
assessment of processes and decisions which are likely to materially
impact the validity of statistical analyses based on a given web-scraped
dataset. It also provides a foundation for multidisciplinary dialogue
between statisticians, domain experts, and programmers (i.e., data
engineers) constructing, validating, documenting, and sharing
web-response datasets. The framework consists of six processes organised
into four stages, which may be examined sequentially and/or iteratively
to document the provenance and quality of new web-scraped datasets,
and/or to investigate already collected web-scraped datasets.

Figure~\ref{fig-framework-overview} illustrates the different stages of
the framework. Beginning with the planning stage, research aims or
analysis questions inform the subsequent retrieval of web-response
objects. After retrieval, the web-response objects need to be
investigated and transformed for analysis usage, often in an iterative
process with multiple evaluations against the desired dataset. Once the
final dataset has been produced and evaluated to be suitable for
analysis, the properties and provenance of the final dataset should be
summarised and documented. Naturally, the materiality of different
dataset properties and provenance information also depends on the
intended research and analysis context.

The framework can also inform the creation of appropriate provenance
artefacts for projects collecting or using web-response datasets. Each
stage involves different combinations of expertise and resources from
data engineering and web-crawling skills to survey methodology and
sampling theory, as well as application and domain-specific knowledge.
In terms of designing provenance artefacts, as mentioned in
Section~\ref{sec-background-provenanceMetadata}, data and workflow
provenance are particularly relevant in the retrieval and transformation
processes, whilst rationale and/or insight provenance are more important
for planning, investigation, and evaluation.

We discuss each stage in turn and outline relevant statistical and/or
technical considerations, common tasks, and suitable documentation
formats and best practices.

\subsection{Plan}\label{sec-framework-planning}

Broadly speaking, the goals of this stage are to conceptualise ideal and
feasible datasets based on the research context and empirical questions.
Ideally, executing an appropriately designed data collection plan would
yield informative data useful for a given inferential hypothesis,
descriptive analysis, or prediction goal. From a statistical
perspective, such data should have a sufficiently strong signal-to-noise
ratio, with relevant variance and well-behaved noise to provide
meaningful insight on stated empirical goals, provided suitable
modelling, estimation, or inference methods are available to extract
those insights. Aligning data with research questions requires careful
definition of the population being studied, attributes of interest, and
meaningful hypotheses about what can be learnt from that data.
Unfortunately, such ideal alignment between data and question becomes
increasingly less likely for web-response datasets as the scale and
complexity of data collection grow. As discussed in
Section~\ref{sec-background-stats}, the volatility and opaqueness of
internet data sources can make it difficult to determine what proportion
of a target population an index list collected via web crawling covers
(i.e., due to the possibility of unindexed units), or to detect factors
that affect the reliability and interpretation of retrieved data (e.g.,
personalisation).

It is unlikely that for a given research context, a single scrape will
provide sufficiently extensive and high-quality data to construct `ideal
datasets'. For example, as we will discuss in
Section~\ref{sec-case-study}, analysing product markets generally
requires collecting data from multiple retailers, all of which likely
use different data structures and website designs. As such, it is useful
to also consider technical, legal, and ethical feasibility in this
stage. In practice, this involves activities such as examining websites
or APIs of a given internet data source, understanding the terms of use,
considering how data came to be and who owns it, determining what
variables and units of analysis are available and in what format, as
well as how comparable attributes (e.g., price, product details, etc.)
are between sources. By considering these factors upfront, it is more
likely that the final web-response dataset will justify the engineering
costs of implementing a web-scraper or similar data collection workflow.

As in all projects, the planning stage is a strong determinant of the
overall success of the project. Although considerations of domain
context, statistical validity, and technical feasibility were reviewed
separately above in Section~\ref{sec-background}, they should be
addressed together in the planning stage to ensure coherence between
research questions and the data. Unfortunately, existing discussions and
proposed guidance on using web-scraping for research often only address
two of these three concerns. For instance, guides to web-scraping for
specific application domains often discuss the practical opportunities
and challenges of collecting novel data of domain interest from internet
sources, with limited mention of statistical considerations. On the
other hand, statistical discussions of internet data sources often lack
any direct consideration of the technical details of web-scraping
workflows or guidance on what types of paradata should be collected
alongside web-response data.

In this stage, documentation is likely to be narrative in format, and
should define the key research questions, characterise the relevant
internet data sources based on initial examinations, and outline the
planned dimensions and contents of the web-response datasets.

\subsection{Retrieve}\label{sec-framework-retrieval}

Given an established plan for a web-response dataset, the retrieval
stage is conceptually equivalent to collection and procurement processes
for traditional survey datasets. The key tasks in this stage are
implementing the necessary workflows for indexing and retrieving the
desired web-response objects (i.e., substantive data), along with
tracking and documenting relevant details about the collection process
(i.e., paradata). Tools and approaches for the former were briefly
discussed in Section~\ref{sec-background-webscrapingTools} and
\ref{sec-background-webTech}. Relevant paradata may include request
headers (e.g., timing, agent type), API queries, target index (URLs),
and response status. For a given target data source and response
objects, the programmatic nature of web-scraping and API queries means
that the retrieval processes can also be documented automatically and
independently of the downstream data use context. As such, the key
artefacts and provenance documentation for this stage are the retrieved
response objects (i.e., `raw' data), records of the implemented
workflows (i.e., reproducible code and scripts), as well as procedural
details in the form of relevant paradata (e.g., logs, request metadata,
etc.). Ensuring that relevant information and data are retained in this
stage can help avoid difficulties with validating constructed variables,
tracing the source of particular data points, and other issues in
characterising and evaluating the final web-response dataset.

\subsection{Investigate, Transform, Evaluate}\label{sec-framework-ite}

Once the retrieval stage has been completed, the construction and
preparation of the final analysis-ready dataset can begin. The
`Investigate, Transform, Evaluate' stage is an iterative process
involving exploratory data analysis, data engineering and wrangling, and
data quality assessment aimed at obtaining a final web-response dataset
suitable for analysing the intended research questions. The three
component tasks in this stage are often referred to collectively using
terms such as `data cleaning' or `data preprocessing'. However, each of
these tasks affects downstream analysis and provenance documentation in
distinct ways. Furthermore, the decision scope of this stage is larger
than for more typical `cleaning' of existing datasets. Addressing each
task separately within a larger iterative process allows for more
precise identification and discussion of the statistical, technical, and
procedural considerations associated with each task, and suitable
accompanying provenance artefacts.

\subsubsection{Investigate}\label{investigate}

The investigation task primarily involves examining and understanding
the retrieved objects, assessing the completeness and quality of
information in the objects, as well as identifying information that can
be used to construct analysis variables (e.g., key-value pairs with
price information). The specific tools and workflows used for this task
can be highly idiosyncratic and depend largely on the format of the
retrieved objects (e.g., media files vs.~JSON objects). As a result,
organising and documenting provenance information and artefacts for this
stage can be quite challenging. Ideally, key assumptions or observations
about the structure or contents of the retrieved objects (i.e., `object
schema' ) should be clearly documented. For structure, assumptions about
object schema could include describing object dimensions or identifying
object elements that fall within the same conceptual variable (i.e.,
equivalent name fields in two different object schemas). For contents,
observations might include counts of missing values, consistency of
values between objects (e.g., set enumeration), or hypothesised
restrictions on possible values for particular elements of the retrieved
objects (e.g., variable type or enumerations).

Unfortunately, as discussed in Section~\ref{sec-background-retrieval},
key assumptions or observations are often lost in all-in-one approaches
to implementing web-scraping workflows. In particular, although the
schema of retrieved objects must be investigated in order to
successfully extract and parse information into structured formats, the
specific assumptions and rationale developed during those investigations
are generally not explicitly documented. At best, assumptions and
observations about the internet data source upon which retrieval and
parsing are based can be deduced from the retrieval or parsing code.
Unfortunately, without explicit documentation, relevant details are
likely to be lost. For example, parsing code which removes white spaces
suggests that strings extracted from response objects had superfluous
white space. However, there is no way to calculate how prevalent this
issue was in the retrieved data based on the parsed data and code alone,
and also no way to reparse the data if it turns out the white space was,
in fact, meaningful to the interpretation of certain observations in the
dataset.

In addition to examining the retrieved response objects, the associated
paradata should also be investigated for anomalous patterns that might
indicate issues that might impact the accuracy and validity of
downstream statistical analyses. Given the exploratory nature of
investigation, this task is least suitable for rigid documentation
protocols. Instead, code notebooks and more narratively flexible forms
of documentation are most recommended. Where possible, it is also useful
to formalise observations into tests that can be used to check whether a
given response object (e.g., collected at a later date) satisfies the
assumptions used in subsequent transformation operations.

\subsubsection{Transform}\label{transform}

Transformation involves implementing data extraction, wrangling, and
cleaning operations based on the observations and assumptions developed
through prior investigations. The workflows in this stage are focused on
extracting, reshaping, and augmenting the retrieved data. As with any
data wrangling workflow, it is useful to separate structural operations
from value modifications. Structural operations such as pivoting,
reshaping, or joining of data (e.g., from multiple online data sources
or existing external datasets) should be justified by prior observations
and assumptions about object schema, and generally can be documented
with code artefacts directly. By contrast, value modifications such as
string parsing and manipulation, or imputation of missing values, will
likely be guided by additional investigations after initial structural
transformations and should ideally be implemented separately after
structural transformation.

Similar to the retrieve stage, the artefacts for this stage can be split
into substantive data (i.e., the transformed data), paradata about the
preparation of that data, as well as metadata summarising the contents
and structure of the substantive data. Paradata for this task might
include descriptive statistics for the retrieved objects (e.g., missing
value counts), data wrangling scripts, as well as details of cleaning
strategies and manipulations.

\subsubsection{Evaluate}\label{evaluate}

Evaluation involves revisiting the statistical, technical, and domain
considerations from the planning stage and comparing the transformed
web-response dataset against the intended research questions. This
includes comparing the overlap in variables constructed from the
web-response objects with those required or desired for use in planned
analyses, as well as assessing the quality (e.g., completeness,
variance, accuracy) of constructed variables. Based on these
assessments, there may be an opportunity or need to adjust the research
aims or questions to better match the available dataset, augment and
transform the constructed dataset further with external datasets or by
extracting additional variables from the response objects, or consider
alternative data sources entirely.

Similar to the planning stage, evaluation requires a combination of
statistical and programming expertise, as well as extensive discussion
with the end users of the data (i.e., domain experts). As such,
documentation of this task will likely also be predominantly narrative
in format, focusing on the rationale for key decisions. Such
documentation will form an important component of the provenance
metadata attached to the final dataset.

\subsection{Summarise}\label{sec-framework-summarise}

The summarise stage begins once a decision has been made to finalise the
web-response dataset for downstream analysis or broader publication and
dissemination. The key consideration for this stage is the appropriate
packaging and summary documentation of the dataset for further use.
Packaging generally includes decisions about distributing the datasets
(e.g., hosting, file formats, etc.), and is often complicated by the
volume of substantive data (e.g., text corpora, extensive time-series).
Summary documentation should at minimum consist of structured,
machine-readable metadata, as well as paradata and code artefacts.
However, for complex datasets, it may also be useful to prepare extended
documentation and commentary on the data preparation process in the form
of dataset manuals, \emph{data papers} {[}31{]}, or other similar report
formats.

\section{Case Study: Online Prices and Product Information for Alcoholic
Beverages in Australia}\label{sec-case-study}

Beyond supporting the planning and production of web-response datasets,
the PRITES framework can also be used to investigate and assess existing
web-response datasets. The framework was developed in part to address
the need to organise and document related work by the authors. This
related work involved adapting a database of web-scraped retail alcohol
prices provided by a commercial retail analytics firm into a research
quality dataset. The data provided by the commercial firm (the `industry
partner') includes price data for products sold online by key alcohol
retailers in Australia and is updated daily. The data were obtained in
order to investigate and analyse the relationship between alcohol harms
and prices, and support the monitoring of pricing patterns in the
Australian retail alcohol market for policy research purposes. The
PRITES framework was used to guide the characterisation and validation
of the provided data as documented in a related `data description paper'
{[}32{]}. In the following sections, we highlight selected applications
of the different PRITES stages and tasks during the preparation of the
provided alcohol prices database. The selected discussions illustrate
how the PRITES framework guided decisions about the data preparation and
downstream analysis.

\subsection{Planning}\label{sec-casestudy-plan}

As discussed in Section~\ref{sec-framework-planning}, a common pitfall
in planning the collection or usage of web-response datasets is failing
to address and integrate relevant statistical, technical or domain
considerations. Initially, the sheer volume and relative novelty of the
provided data led to an extremely ambitious and broad research agenda.
In particular, the scale and extent of the commercial web-scraping
operation appeared to offer far more extensive product and retailer
coverage compared to data sources previously used by the public health
domain experts in the team (e.g., manual in-store audits, self-reported
consumer surveys). Furthermore, the data appeared to include an
extensive selection of auxiliary variables (e.g., alcohol percentages,
packaging types, package size, country of origin etc.), and were already
harmonised to match identical products across retailers to allow for
product level analysis. However, the content, completeness and accuracy
of the additional variables and harmonisation matches were not
necessarily suitable for some of the intended research questions from
the perspective of statistical validity.

In order to assist the domain experts reason about the likely quality of
variables constructed from the provided data for particular research
questions, we adapted the concept of ``shoe-leather costs'' of data
collection {[}33{]}. Team members were asked to consider how salient a
particular product attribute is on an in-store shelf and/or the product
packaging itself as a proxy for the likelihood that the web-scraped data
contained sufficiently high-quality information to construct a
corresponding analysis variable, as well as for the likely `cost' of
extracting that information (e.g., text from product description
vs.~standardised volume measures). This exercise helped to focus the
research direction towards using variables that were more likely to be
complete and accurate.

\subsection{Retrieval}\label{sec-casestudy-retrieve}

Our understanding of this stage was predominantly informed by
discussions with the data engineers who built, operated and maintained
the web-scraping infrastructure within the industry partner firm. The
framework helped us to identify and pose data quality questions in terms
of data engineering workflow decisions such as when the web-scrapers
were run each day, how failed retrieval attempts were handled, and
whether they had observed any hostile `anti-scraping' behaviour from
particular retailer websites. In the absence of extensive paradata, the
discussions helped us understand the collection process and enumerate
possible sources of statistical bias and errors as well as drivers of
data quality not present in academic web-scraping projects. For example,
troubleshooting and fixes for failed scrapes are generally implemented
within hours by the industry partner's full-time data engineers, due to
the real-time price monitoring offered to their clients. By contrast,
web-scraping projects in academic settings are conducted with far fewer
resources, and unlikely to be monitored full-time. Without full-time
monitoring and timely fixes, it is more likely that collected data will
be incomplete and have irregular time intervals between observations
complicating comparisons across time (e.g., online prices could differ
based on time of day).

\subsection{Investigation, transformation and
evaluation}\label{sec-casestudy-ite}

Investigation of the provided database required extensive data
manipulation and descriptive analysis, and eventually led to the
collection of an additional web-scraped dataset in order to augment and
validate the provided price and product data. There was considerable
effort required to understand the semantics of the variables provided in
the database, as well as the comparability of certain variables (e.g.,
product category, promotions) across retailers. This included comparing
product listings and webpages from the retailer websites against the
provided variables, and seeking clarification and confirmation from the
industry partner on the interpretation of certain variables. We were
also able to obtain some intermediate unparsed `raw' tabular content
extracted from the scraped product pages. However, the raw HTML response
objects were not retained, making it difficult to fully reconstruct the
rationale behind the final transformed data as discussed in
Section~\ref{sec-framework-ite}.

We engaged in multiple investigation and evaluation loops to adjust the
target research questions. For example, the presence of
\texttt{price\_was}, \texttt{price\_now} and \texttt{promo} variables in
the industry dataset generated some initial interest from collaborators
to examine patterns of discounting across different retailers. However,
upon further discussion and investigation, it became clear that data
only captured some discounting formats. In particular, only actively
advertised price-based discounts would reflect in the
\texttt{price\_was} and \texttt{price\_now} variables (e.g., 10\% off),
whilst ``indirect'' discounting through bulk-buy promotions (e.g.,
2-for-1), or discounts applied in the shopping cart would not be
reflected.

Another important investigation was characterising the quality of
variables to do with the alcohol concentration, package volume and
number of standard drinks for a given product. These variables are
required to analyse `minimum unit price' policies. Minimum unit price
policies prescribe limits on the lowest allowable `price per standard
drink' and have been investigated in previous work by members of the
research team {[}34{]}. Unfortunately, we discovered numerous data
issues including missing data and implausible values (e.g., impossible
bottle volumes, string values rather than percentages etc.). This led us
to conduct a secondary web-scrape in an attempt to collect more accurate
product information, and to explore the use of large-language models to
suggest corrections to implausible values and other manual data entry
errors. These secondary augmentations were all documented as
`transformation' steps in the dataset description.

\subsection{Summary}\label{sec-casestudy-summary}

The notion of a `finalised dataset' is difficult to apply to a database
of price data that is updated on a daily basis. However, the specific
research analyses conducted on the provided data ultimately examine a
time-limited subset of the available data. Furthermore, additional
product information has been used to filter and limit analysis to
specific subsets of products based on features such as popularity or
product type (e.g., wines vs.~spirits). The combination of time and
product sub-setting for specific analyses gives rise to a collection of
derived `finalised datasets' corresponding to different research
questions or hypotheses.

We summarise and describe the provenance, properties and limitations of
this family of datasets in a related data description paper by the
co-authors. Instead of providing a chronological recount of the data
preparation process, we used the PRITES framework to highlight decisions
and properties material for assessing the suitability and validity of
the derived datasets in subsequent empirical analyses.

\section{Discussion}\label{sec-discussion}

\subsection{Limitations}\label{limitations}

Compared to existing templates and guides for documenting datasets and
conducting web-scraping projects, our framework lacks specific guidance
on how to collect, process or analyse web-response data. However, we
intend for the framework to serve as a supporting structure for
identifying, discussing and incorporating multidisciplinary
considerations and expertise throughout the conception to completion of
empirical analyses based on web-response datasets. Furthermore, it would
not be practical or feasible to fully enumerate the many statistical,
technical and domain-specific considerations relevant to any given
internet data source or web-response dataset.

\subsection{Future Work}\label{future-work}

The development of the PRITES framework and the associated example case
study also reveal opportunities to develop web-scraping tools that
support the collection of relevant paradata, documentation of the
characterisation process and findings, and alternative transformation
choices. Integrating paradata collection into the retrieval,
investigation and transformation stages could reduce the risk of
preventable data quality loss associated with procedural and/or
documentation issues and support the application of relevant statistical
expertise in projects using web-response datasets. As discussed
previously in Section~\ref{sec-framework-retrieval}, many forms of
paradata in the retrieval stage could be collected automatically. The
identification of key paradata, developing tooling to collect that
information, and the mapping of those paradata to statistical
considerations are all avenues for future work.

The framework also highlights three relevant documentation formats:
narrative, code and structured metadata. For each of these formats,
there are opportunities to develop tools specific to the production of
web-response datasets. For example, when investigating retrieved HTML
pages, it could be useful to create modified versions of the HTML pages
to serve as `visual data dictionaries'. Rather than simply describing in
text where information (e.g., prices) was extracted from, a `visual data
dictionary' could highlight the relevant HTML elements on the source
webpage and show the names of variables in the final dataset constructed
from information in those elements. Finally, when considering
transformation choices, given the overlap with general data wrangling
workflows, well-commented code scripts and replication packages are
likely to be sufficient in most cases.

\section{Conclusion}\label{conclusion}

This work attempts to integrate statistical, technical and domain
considerations in the preparation and assessment of datasets constructed
using data from the World Wide Web. We propose the term `web-response
datasets' to refer to such datasets, and `internet data sources' to
refer to the original source of information in those datasets. We review
existing programmatic tools and approaches to conducting web-scraping
tasks and collecting web-response objects, and connect the all-in-one
design of such tools with challenges in documenting and interpreting the
provenance and statistical properties of web-response datasets. We also
briefly review potential issues with the statistical validity of
analyses which use web-response datasets, and existing frameworks for
assessing the quality and suitability of such data.

Based on the existing literature and a related case study, we offer the
PRITES framework for conceptualising the production process of
constructing web-response datasets. The framework organises six related
web-scraping, data wrangling and quality assessment tasks into four
workflow stages along with recommended provenance documentation for each
task. It can be operationalised both as a workflow for new web-scraping
projects, as well as an assessment tool existing web-scraped datasets.
We illustrate the latter in the included case study, and show how the
framework can guide the interdisciplinary dialogue required to reconcile
technical success in retrieving web-response objects and the
construction of analysis-ready datasets suitable for broader research
goals. Finally, we discuss limitations of the framework compared to more
detailed guides, and how the framework might inform the design of
web-scraping tools and documentation artefacts that facilitate smoother
workflows for researchers working with web-response datasets.

\subsection*{References}\label{references}
\addcontentsline{toc}{subsection}{References}

\phantomsection\label{refs}

\end{document}